\documentclass{iopconfser}

\usepackage{a4wide,amssymb,cite,graphicx}
\usepackage{epsfig}
\usepackage{hyperref}

\def\beq{\begin{equation}}
\def\eeq{\end{equation}}

\begin{document}

\noindent Conference Proceedings for BCVSPIN 2024: Particle Physics and Cosmology in the Himalayas\\Kathmandu, Nepal, December 9-13, 2024 

\title{Cosmic Colliders: \\ High Energy Physics with First-Order Phase Transitions}

\author{Bibhushan Shakya}

\affil{Deutsches Elektronen-Synchrotron DESY, Notkestr.\,85, 22607 Hamburg, Germany}

\email{bibhushan.shakya@desy.de}
~
\begin{abstract}
Collisions of vacuum bubbles in the early Universe can act as cosmic-scale high-energy colliders with energy reach close to the Planck scale. Such  ``cosmic colliders" would represent the most energetic phenomena in our cosmic history, transcending any temperature or energy scale ever reached in our Universe, opening tremendous opportunities for particle physics and cosmology. Such configurations are realized during first-order phase transitions with runaway bubbles -- a topic of significant current research interest as a promising cosmological source of gravitational waves. We discuss recent developments and challenges in the physics of such cosmic colliders, as well as their broad applications for particle physics and cosmology, from dark matter to leptogenesis to gravitational waves.

\end{abstract}

\section{Introduction}

First order phase transitions (FOPTs) \cite{Hogan:1983ixn, Witten1984, Hogan:1986qda, KosowskyTurnerWatkins-GW-92, kosowsky92-1, Kamionkowski_1994}, which proceed via the nucleation of bubbles of true vacuum in a false/metastable vacuum background, are well-motivated in many Beyond the Standard Model (BSM) scenarios, and have received significant interest and attention in recent years as one of the most promising cosmological sources of stochastic gravitational waves (GWs) \cite{Grojean:2006bp,Caprini:2015zlo,Caprini:2018mtu,Caprini:2019egz,Athron:2023xlk}. 

 It has now been clearly established, through a wide variety of analytic as well as numerical studies, that some FOPT configurations can attain the so-called runaway behavior, where the vacuum energy released during the phase transition can become concentrated on the bubble walls that separate the two phases, boosting these bubble walls to ultra-relativistic speeds. This occurs when friction effects from the plasma on the bubble walls are not significant, which is the case, e.g.\,for supercooled phase transitions, transitions in cold sectors, or sectors without gauge bosons  (see \cite{Giudice:2024tcp} for a broader discussion). In the context of GWs, these are the scenarios where the GW signal is dominated by the scalar field energy densities in the bubble walls after collision, which has been extensively studied in the literature\,\cite{Kosowsky:1991ua,Kosowsky:1992rz,Kosowsky:1992vn,Kamionkowski:1993fg,Caprini:2007xq,Huber:2008hg,Bodeker:2009qy,Jinno:2016vai,Jinno:2017fby,Konstandin:2017sat,Cutting:2018tjt,Cutting:2020nla,Lewicki:2020azd}.

 However, the potential for the collisions of such ultrarelativistic bubbles to act as high energy ``cosmic colliders"\footnote{Unrelated to the similar-sounding cosmological collider physics \cite{Arkani-Hamed:2015bza} program.} capable of producing particles with energies or masses far above the temperature of the thermal plasma has remained largely unappreciated over the years. Most studies of FOPTs study bubble dynamics and energy dissipation at a classical level (e.g. \cite{Hawking:1982ga,Kleban:2011pg,Chang:2008gj}), and  focus on dynamics at lengths scales corresponding to the typical size of bubbles, since this is the scale relevant for GW production, rather than the (many orders of magnitude smaller) length scale corresponding to the boosted bubble wall thickness. Particle production, on the other hand, is an inherently quantum phenomenon, and requires a quantum-field-theoretical approach. The simplest ``quantum" treatment of bubble collisions \cite{Jinno:2017fby}, approximating the ultrarelativistic bubble walls as condensates of quanta of the background field moving at ultrarelativistic velocities, also concluded that such collisions are extremely inefficient at producing high energy particles, since the associated interaction cross-section gets parametrically suppressed at high energies. However, this simplistic approach is also inconsistent with the known classical evolution of the background field at collision, and therefore does not fully capture the underlying physics.

What is needed is a more rigorous approach that takes into account both the classical evolution of the background field during the bubble collision process and the quantum nature of the particle production process.  This was done in \cite{Watkins:1991zt}, where particle production from colliding bubbles was calculated using the effective action formalism. This formalism was subsequently explored numerically in \cite{Konstandin:2011ds} in the context of cold baryogenesis, and further developed in \cite{Falkowski:2012fb} with semi-analytic approaches. These studies established that particle production from bubble collisions can indeed be an efficient process that can convert a significant fraction of the vacuum energy released during the course of the transition into energetic particles. Efficient production of particles with energies or masses several orders above the the scale of the phase transition, though, was found to occur only in special configurations (the so-called elastic collisions, where bubble walls bounce back after colliding) rather than being a general phenomenon. However, more recent studies have demonstrated, through heuristic and analytic arguments \cite{Shakya:2023kjf} as well as numerical studies \cite{Mansour:2023fwj}, that the efficiency for producing high-energy phenomena from bubble collisions follows a universal power law, which is independent of the finer details of the collision process (such as whether the collision is elastic or inelastic) as long as the bubble walls are ultra-relativistic. Therefore, the production of particles with very high masses or energies --  the realization of a cosmic collider -- in an inevitable phenomenon in any FOPT with runaway bubbles.

\section{Formalism}

Here we provide a brief outline of the formalism to calculate particle production from bubble collisions; the interested reader is referred to \cite{Watkins:1991zt,Falkowski:2012fb,Mansour:2023fwj,Shakya:2023kjf} for greater details. Consider a FOPT associated with a scalar field $\phi$ with vevs $\langle \phi\rangle =0,v_\phi$ in the false (unbroken) and true (broken) vacua, where we parameterize the energy density between the two vacua as
\beq
\Delta V \equiv V_{\langle \phi \rangle=0}-V_{\langle \phi \rangle=v_\phi}=c_V\,v_\phi^4\,.
\eeq

The probability of particle production from the dynamics of the field $\phi$ can be calculated as the imaginary part of its effective action \cite{Watkins:1991zt}, 
\beq
\mathcal{P}=2 \,\mathrm{Im}\,(\,\Gamma[\phi\,]\,),
\eeq
where the effective action $\Gamma[\phi\,]$ is the generating functional of one-particle irreducible (1PI) Green
functions  
\beq
\Gamma[\phi\,]=\sum_{n=2}^\infty \frac{1}{n !}\int d^4 x_1 ... d^4 x_n \Gamma^{(n)}( x_1,...,x_n)\phi (x_1)...\phi(x_n).
\label{expansion}
\eeq
The leading ($n=2$) term gives
\beq
\mathrm{Im}\,(\Gamma[\phi])=\frac{1}{2}\int d^4x_1 d^4 x_2 \phi(x_1)\phi(x_2) \int \frac{d^4 p}{(2\pi)^4}e^{i p (x_1-x_2)} \mathrm{Im}(\tilde{\Gamma}^{(2)}(p^2))\,,
\eeq
where $\tilde{\Gamma}^{(2)}$ is the Fourier transform of $\Gamma^{(2)}$. Assuming a collision between two planar walls of equal thickness  in the $z-$direction, the number of particles produced per unit area of colliding bubble walls can be written as \cite{Watkins:1991zt,Falkowski:2012fb}
\beq
\frac{N}{A}= 2 \int\frac{dp_z\,d\omega}{(2\pi)^2}\,|\tilde{\phi}(p_z,\omega)|^2 \,\mathrm{Im}[\tilde{\Gamma}^{(2)}(\omega^2-p_z^2)]\,.
\label{interpretation}
\eeq
The background field at collision can thus be interpreted as a collection of Fourier modes with definite energies. These Fourier modes correspond to off-shell excitations $\phi^*_p$ of the background field that carry definite four-momenta $p^2=\omega^2-p_z^2>0$, and can decay into particles that couple to the background field. 

Following a change of variables, the above formula can be simplified as \cite{Falkowski:2012fb} 
 \beq
\frac{N}{A}=\frac{1}{2 \pi^2}\int_{p_{\mathrm{min}}^2}^{p_{\mathrm{max}}^2} d p^2\,f(p^2) \,\mathrm{Im} [\tilde{\Gamma}^{(2)}(p^2)].
\label{number}
\eeq
Here $f(p^2)$ incorporates the details and nature of the collision process, i.e. the spacetime dynamics of the background field configuration. The lower limit $p_{min}=2\,m$ (for pair production), is set by the mass of the particle species being produced, or the inverse size of the bubbles at collision, (at lower momenta, the existence of multiple bubbles needs to be taken into account), whichever is greater. The upper cutoff is provided by $p_{max}=2/l_w=2\gamma_w/l_{w0}$, the inverse boosted thickness of the bubble walls at collision. This represents the physical length (equivalently, energy) scale that is probed by the collision process, and it should therefore be possible to produce particles with masses or energies up to this scale. 

The bubble wall thickness at nucleation is $l_{w0}\sim v_\phi^{-1}$. In the runaway regime, the boost factor of the bubble wall is known to grow linearly with the size of the bubble, $\gamma\sim R/R_0$, where $R (R_0)$ is the bubble radius (at nucleation). This can be understood from simple energy conservation arguments: since the latent vacuum energy released from the volume of the bubble ($\sim R^3$) is being concentrated on the surface of the bubble, i.e. the bubble walls ($\sim R^2$), the energy per unit surface area must grow linearly. Now, for the bubbles to percolate so that the phase transition can complete and the true vacuum can be established everywhere, there must be at least one bubble per Hubble volume, i.e. a typical bubble radius at collision, $R_*$, must be smaller than the Hubble radius, $R_* H < 1$, where $H$ is the Hubble parameter.  Assuming the total energy density in the Universe is of the same order as the latent vacuum energy, $H^2= \frac{8\pi\Delta V}{3 M_P^2}=\frac{8\pi c_V v_\phi^4}{3 M_P^2}$, where $M_{P}$ is the Planck mass.  From these, and using $R_0\sim v_\phi^{-1}$, we have
\beq
p_{max}=2\gamma_w/l_{w0}=\frac{2R_*}{R_0}v_\phi < \frac{2v_\phi^2}{H}\sim M_P
\eeq
up to $\mathcal{O}(1)$ factors. Therefore, the bubble collisions can indeed approach Planck scale energies! 

In practice, the duration of the FOPT, parameterized as $\beta$, is shorter than a Hubble time; generally $\beta/H\sim \mathcal{O}(10-10,000)$. Consequently,  $p_{max}\lesssim M_P/(\beta/H)$. Nevertheless, note that this is independent of the scale of the phase transition $v_\phi$, and can be many orders of magnitude larger.  

Let us now see how efficient the collisions are at probing such high energy scales. In Eq.\,\ref{number}, the $f(p^2)$ term represents the efficiency factor for producing background field excitations at a given scale $p$. The component that accounts for the collision of ultrarelativistic bubbles takes a universal form for $p\gg v_\phi$
\cite{Watkins:1991zt,Falkowski:2012fb,Mansour:2023fwj,Shakya:2023kjf}
\beq
f(p^2)=\frac{16 v_{\phi}^2}{p^4}\, \mathrm{Log}\left[\frac{2(1/l_w)^2-p^2+2(1/l_w)\sqrt{(1/l_w)^2-p^2}}{p^2}\right]\,.
\label{eq:felastic}
\eeq
The logarithmic factor is $\mathcal{O}(10)$ over large regions of parameter space of interest. The imaginary part of the 2-point 1PI Green function $\Gamma^{(2)}$ can be calculated using the optical theorem \cite{Watkins:1991zt,Falkowski:2012fb,Giudice:2024tcp}, and depends on the nature of the interaction. It has been found that the above formalism is not gauge-independent \cite{Giudice:2024tcp}, in particular when calculating the production of gauge bosons from bubble collisions; nevertheless, physical results can be extracted in the high energy limit.

\section{BSM Applications and Phenomenology}

~

Here we briefly discuss some applications and phenomenology of cosmic colliders. 

~

\noindent\textbf{Ultraheavy Dark matter} (based on \cite{Giudice:2024tcp})

Ultrarelativistic bubble collisions provide a viable non-thermal mechanism to produce ultraheavy dark matter with mass many orders of magnitude higher than the scale of the phase transition or the temperature of the plasma (for earlier papers with similar ideas, see \cite{Falkowski:2012fb,Freese:2023fcr}).  

\begin{figure}[t]
\begin{center}
\includegraphics[width=0.49\textwidth]{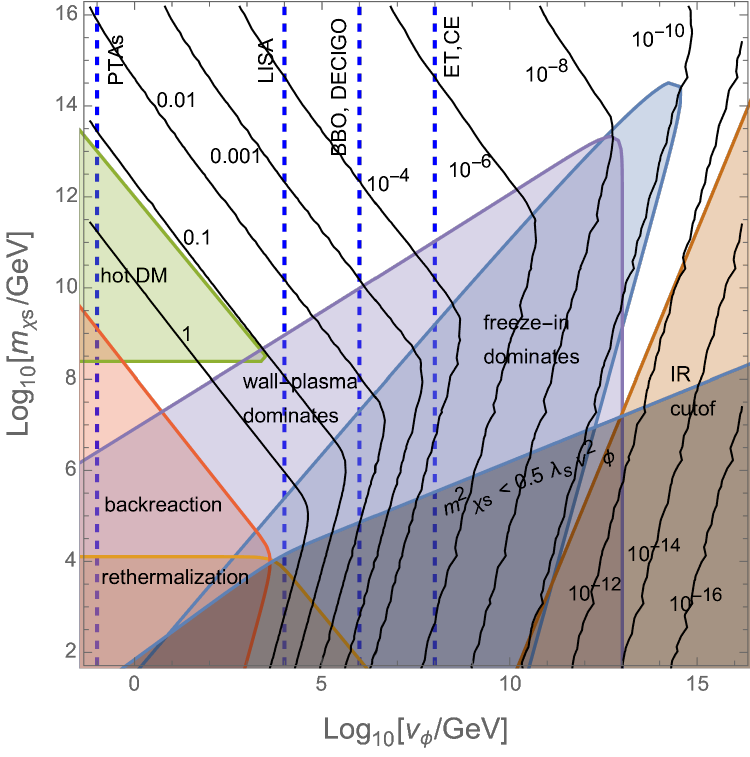}~
\includegraphics[width=0.49\textwidth]{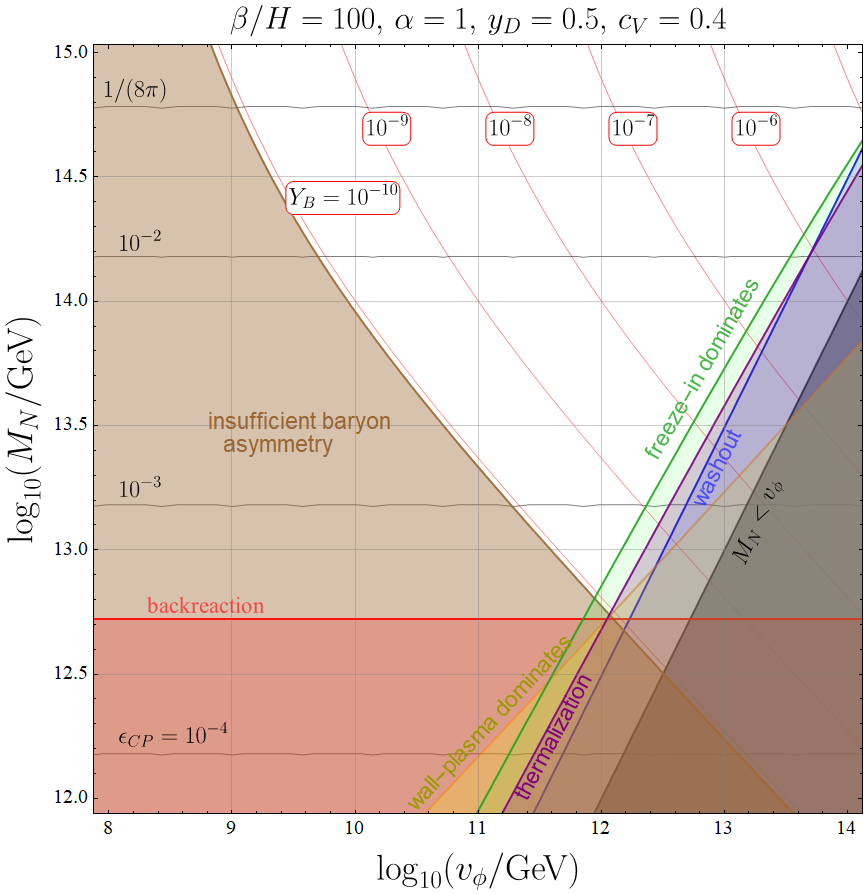}
\end{center}
\caption{
Ultraheavy dark matter (left, from \cite{Giudice:2024tcp}) and nonthermal leptogenesis (right, from \cite{Cataldi:2024pgt}) from ultrarelativistic bubble collisions.  
}
\label{fig:scalarrelic}
\end{figure}

Consider scalar DM $\chi_s$  that couples to the background field $\phi$ via $\frac{1}{4}\lambda_s \phi^2 \chi_s^2$, and can be produced via $\phi^*_p\to\chi_s^2,~ \phi \chi_s^2$. The DM relic abundance from bubble collisions can be calculated as  (see \cite{Giudice:2024tcp} for details)
\beq
\Omega_\chi h^2\approx 0.1\,\frac{\beta/H}{10}\left(\frac{\alpha}{(1+\alpha)g_* c_V}\right)^{1/4}\frac{\lambda_s^2\,m_{\chi_s}\,v_\phi}{(24~\mathrm{TeV})^2}\,\left[\frac{v_\phi^2}{m_{\chi_s}^2}+\frac{1}{16\pi^2}\,\ln\left(\frac{2\,\gamma_w/l_{w0}}{(2m_{\chi_s}+m_\phi)}\right)\right]\,.
\label{eq:scalarabundance}
\eeq
The parameter space where the correct dark matter relic abundance is realized is shown in the left panel of Fig.\,\ref{fig:scalarrelic}; contours represent the required size of the coupling $\lambda_s$. This shows that DM production from bubble collisions is a viable mechanism over a large region of parameter space spanning many orders of magnitude. For detailed dicussions of the various regions in the plot, see \cite{Giudice:2024tcp}.

~

\noindent\textbf{Nonthermal Leptogenesis}  (based on \cite{Cataldi:2024pgt})

Cosmic colliders also provide novel means to produce the baryon asymmetry of the Universe (for other related ideas, see \cite{Katz_2016}). Consider the standard leptogenesis scenario, where heavy right-handed neutrinos (RHN) $N$ that can explain the tiny SM neutrino masses through the type-I seesaw mechanism, if produced in the early Universe, can decay to produce a lepton asymmetry that is converted to a baryon asymmetry by sphaleron processes. For type-I seesaw with $\mathcal{O}(1)$ couplings, the RHNs are extremely heavy, $M_N\sim 10^{14}$ GeV. In this regime, washout processes are known to be active, and there are no observable aspects of such high-scale leptogenesis.  

 We can mirror the RHN interactions with the SM into a hidden sector undergoing a FOPT. Consider a scalar $\phi$ that undergoes the FOPT, and a dark sector fermion $\chi$ such that $\phi\,\chi$ is a gauge singlet under the symmetry broken by the $\phi$ vev (analogous to the $Lh$ combination in the SM). This enables us to write the following Lagrangian terms for the RHN (see e.g.\,\cite{Roland:2014vba,Shakya:2018qzg,Roland:2016gli,Shakya:2016oxf,Roland:2015yoa,Shakya:2015xnx,Morrison:2022zwk} for details and various applications of such neutrino portal frameworks): 
\beq
\mathcal{L}\supset y_\nu \overline{L} h N+y_D \chi \phi N +M_N \overline{N}^c N\,.
\eeq
The first term is the standard SM Dirac neutrino mass, and the second term is its dark sector analogue that couples the RHN to dark sector states.  The collision of ultrarelatvistic bubble walls can produce RHNs via $\phi_p^*\to \chi N$ even when $M_N\gg v_\phi, T$, where $T$ is the temperature of the bath. The parameter space where this configuration can produce the observed baryon asymmetry is shown in the right panel of Fig.\,\ref{fig:scalarrelic}. Note that the washout processes that can erase the produced asymmetry are inactive in this case. For detailed dicussions of the various regions in the plot, see \cite{Cataldi:2024pgt}.

~

\noindent\textbf{Gravitational Waves} (based on \cite{newGW})

\begin{figure}[t]
\begin{center}
\includegraphics[width=0.6\textwidth]{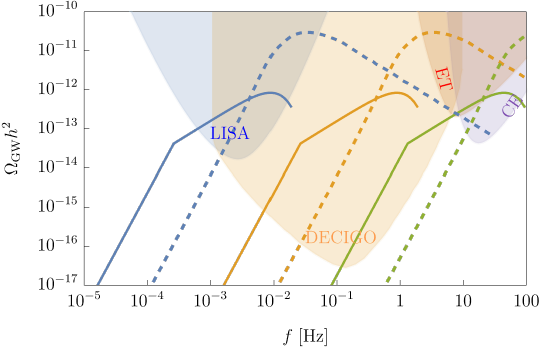}
\end{center}
\caption{
Gravitational wave signals from relativistic particles produced by cosmic colliders (solid curves). Dashed curves denote the standard GW signals from the scalar field energy densities in the bubble walls.  Plot from \cite{newGW}.
}
\label{fig:GW}
\end{figure}

Understanding of the physics of ultrarelatvistic bubble collisions is relevant not only for high energy and particle physics applications, but also for understanding the gravitational wave signals generated by FOPTs. The efficient production of a population of relativistic particles from bubble collisions can create a new source of GWs \cite{newGW} beyond the traditional ones (bubble walls, sound waves and turbulence (see references in \cite{Caprini:2019egz}), feebly interacting particles \cite{Jinno:2022fom}) associated with FOPTs. The GW signal produced from such relativistic particle configurations is plotted in Fig.\,\ref{fig:GW} (for a detailed discussion, see \cite{newGW}).  While this new contribution (solid curves) has a smaller amplitude than the signal arising from the scalar field energy densities during collision (dashed curves), it can dominate the signal in the infrared since the particles produced from bubble collisions survive long after the bubbles have disappeared, producing GWs for a longer time at lower frequencies. This creates a distinct shift of the spectral slope of the GW signal from a cubic to a linear falloff around $k/\beta \approx 0.1$, which could provide an observable feature at upcoming GW detectors.

\bibliographystyle{jhep}
\bibliography{Ref}{}

\end{document}